\documentclass[aps,prl,twocolumn,preprintnumbers,superscriptaddress,10pt]{revtex4-1}
\usepackage[mathscr]{eucal}
\usepackage{epsf,color,amsmath,amssymb,amsfonts}
\usepackage{enumerate}
\usepackage{hyperref}
\usepackage{graphicx}
\usepackage{psfrag}
\usepackage{dcolumn,bm}
\usepackage{MnSymbol}
\usepackage{mdframed}

\newcommand{\bea}{\begin{eqnarray}}
\newcommand{\beal}[1]{\begin{eqnarray}\label{#1}}
\newcommand{\eea}{\end{eqnarray}}
\newcommand{\be}{\begin{equation}}
\newcommand{\bel}[1]{\begin{equation}\label{#1}}
\newcommand{\ee}{\end{equation}}

\newcommand{\bit}{\begin{itemize}}
\newcommand{\eit}{\end{itemize}}
\newcommand{\ben}{\begin{enumerate}}
\newcommand{\een}{\end{enumerate}}

\newcommand{\mt}[1]{\textrm{\tiny #1}}
\newcommand{\nc}{N_\mt{c}}

\newcommand{\eps}{{\cal E}}

\renewcommand{\pl}{{\cal P}_L}
\newcommand{\s}{{\cal S}}
\newcommand{\pt}{{\cal P}_T}

\newcommand{\thy}{t_\mt{hyd}}
\newcommand{\Thy}{T_\mt{hyd}}
\newcommand{\tmax}{t_\mt{max}}
\newcommand{\tlin}{t_\mt{lin}}

\begin{document}
\onecolumngrid

\preprint{ICCUB-13-070,\,ITP-UU-13/11}

\title{
From full stopping to transparency in a holographic model of heavy ion collisions}
\author{Jorge Casalderrey-Solana} 
\affiliation{Departament d'Estructura i Constituents
de la Mat\`eria, 
Universitat de Barcelona, Mart\'\i \ i Franqu\`es 1, 08028 Barcelona, Spain}
\author{Michal P.~Heller}
\altaffiliation[On leave from: ]{\emph{National center for Nuclear Research,  Ho{\.z}a 69, 00-681 Warsaw, Poland.}}
\affiliation{Instituut voor Theoretische Fysica, Universiteit van Amsterdam \\
Science Park 904, 1090 GL Amsterdam, The Netherlands}
\author{David Mateos}
\affiliation{{Instituci\'o Catalana de Recerca i Estudis Avan\c cats (ICREA), 
Barcelona, Spain}}
\affiliation{{Departament de F\'\i sica Fonamental \&  Institut de Ci\`encies del Cosmos (ICC), Universitat de Barcelona, Mart\'{\i}  i Franqu\`es 1, E-08028 Barcelona, Spain}}
\author{Wilke van der Schee}
\affiliation{Institute for Theoretical Physics and Institute for Subatomic Physics,
Utrecht University, Leuvenlaan 4, 3584 CE Utrecht, The Netherlands}


\begin{abstract}
We numerically simulate planar shock wave collisions in anti-de Sitter space as a model for heavy ion collisions of large nuclei. We uncover a cross-over between two different dynamical regimes as a function of the collision energy. At low energies the shocks first stop and then explode in a manner approximately described by hydrodynamics,  in close similarity with the Landau model.
At high  energies the receding fragments move outwards at the speed of light, with  a region of negative energy density and negative longitudinal pressure trailing behind them. The rapidity distribution of the energy density at late times 
around mid-rapidity is not approximately boost-invariant but Gaussian, 
albeit with a width that increases with the collision energy. 
\end{abstract}
\maketitle

\noindent
{{\bf 1. Introduction.}}
Holography 
has provided successful toy models for the study of (near)equilibrium properties of the quark-gluon plasma created in heavy ion collisions (HIC) at RHIC and LHC (see e.g.~\cite{CasalderreySolana:2011us} for reviews). Applying holography to the far-from-equilibrium early stage of a HIC is challenging and interesting. The challenge arises because one must solve Einstein's equations in a dynamical setting, which generically must be done numerically \cite{Chesler:2010bi,Chesler:2008hg}. The interest lies in that understanding the strong coupling limit described by holography  may help us bracket the real-world situation.

Here we will follow the approach of Ref.~\cite{Chesler:2010bi}, in which a HIC was toy-modeled as a collision of two planar shock waves of finite thickness in anti-de Sitter space (AdS). In the dual conformal field theory (CFT) this corresponds to a collision of two infinite sheets of energy characterized by a stress tensor whose only non-zero component is
$T_{\pm \pm }(z_\pm) = \frac{N_{c}^{2}}{2 \pi^{2}} \, 
\rho^4 \, e^{-z_\pm^2/2 w^2}$, 
where $z$ is the `beam direction', $z_\pm = t\pm z$, $w$ is the width of the sheets and the sign depends on the direction of motion of the shock. We choose $t=0$ to correspond to the time at which the two shocks would exactly overlap if there were no interactions. We will work with energy densities, energy fluxes and pressures normalized as $({\cal E}, {\cal S}, {\cal P}_L, {\cal P}_T)=\tfrac{2\pi^2}{\nc^2}
(-T^t_t, T^z_t, T^z_z, T^{x_\perp}_{x_\perp})$.
We will thus refer to $\rho^4$ as the maximum energy density of the incoming shocks, which is related to the energy per unit transverse area $\mu$ used in \cite{Chesler:2010bi} through $\mu^3 = \sqrt{2\pi} \, \rho^4 w$. 
Scale invariance of the CFT implies that the physics only depends on the dimensionless product $\rho w$. Ref.~\cite{Chesler:2010bi} chose $\mu w_\mt{CY} = 0.75$, corresponding to $\rho w_\mt{CY} \simeq 0.64$. Note that for the incoming shocks one has $\eps= \pl = \mp \s$ and $\pt=0$.

Given the simplicity of the model, we will not attempt to match the values of $\rho$ and $w$ to a specific HIC. Instead, we note that, in a real HIC, the product $\rho w$ decreases as $\gamma^{-1/2}$ as the total center-of-mass energy of the collision, $\sqrt{s_\mt{coll}}=2\gamma M_\mt{ion}$, increases. This suggests that HICs at increasingly higher energies may be   modeled by decreasingly smaller values of $\rho w$ \cite{transverse}. We will  therefore simulate collisions with several values of $\rho w$ ranging from $2 \rho w_\mt{CY}$ to 
$\tfrac{1}{8} \rho w_\mt{CY}$. We will refer to the former  as `thick shocks' and to the latter as `thin shocks'. We will focus on our physical results and refer the reader to \cite{Chesler:2010bi} for technical details \cite{regulator}. We will work with fixed $\rho$ and vary $w$, and hence think of low-energy and high-energy collisions as modeled by thick and thin shocks, respectively. 
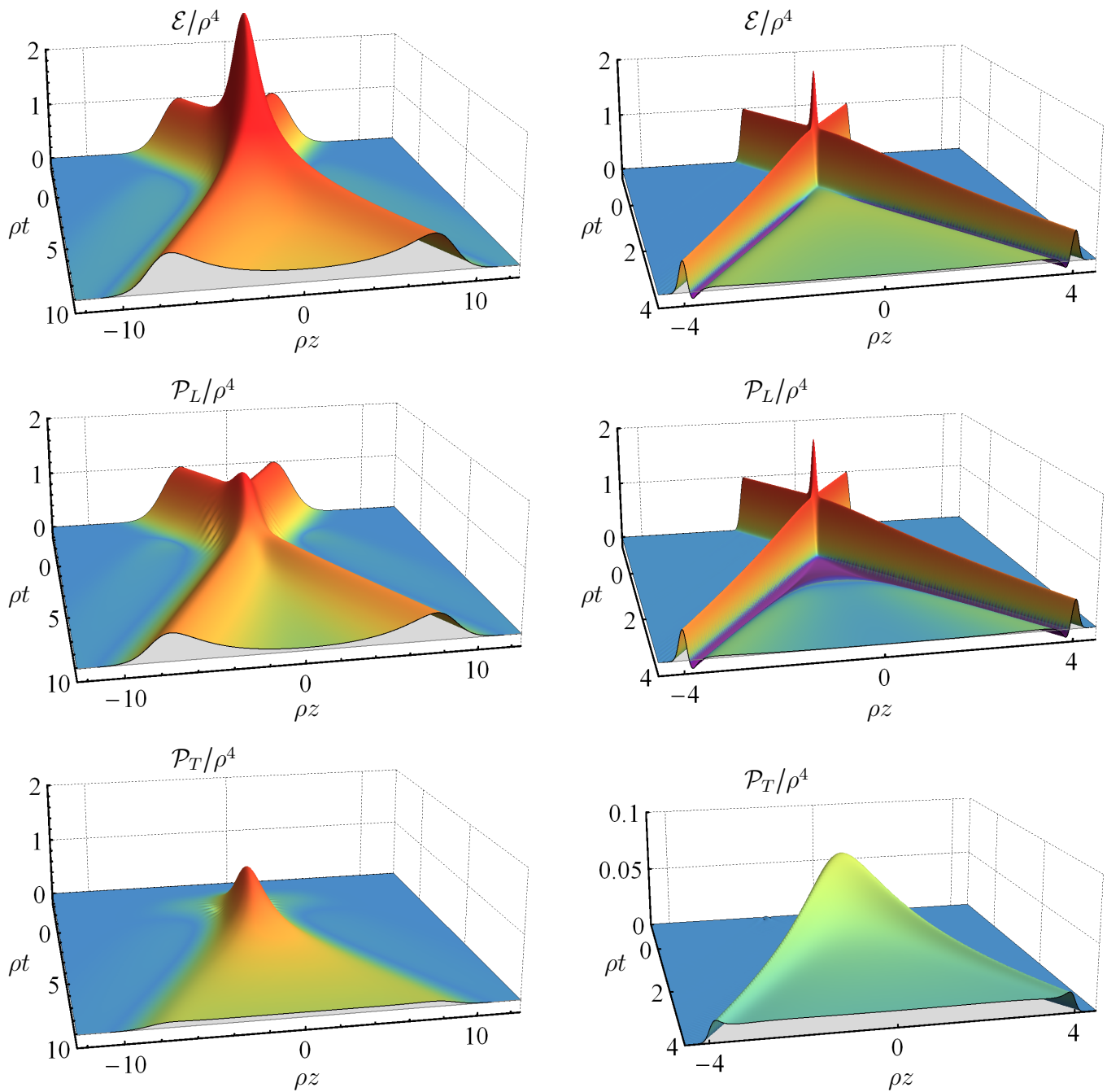
\begin{figure*}
\begin{tabular}{cc}
\includegraphics[width=0.46 \textwidth]{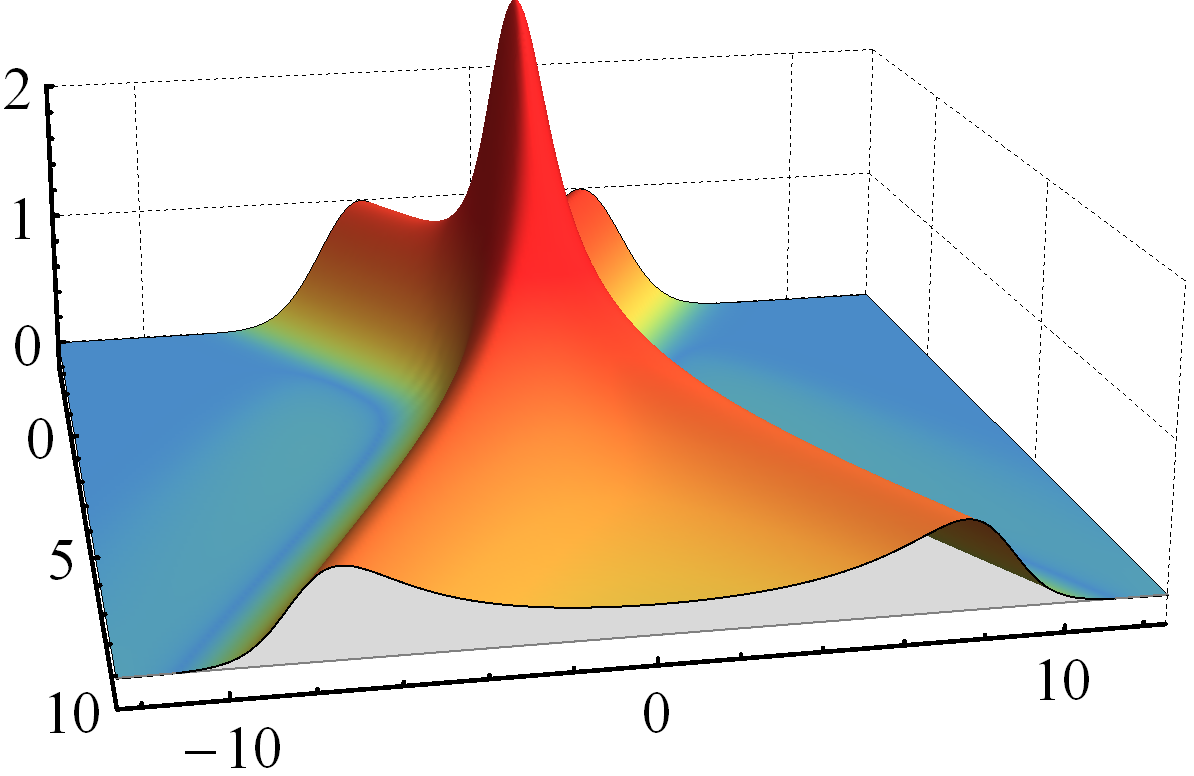}
\put(-167,145){\mbox{\large ${\cal E}/\rho^4$}}
\put(-243,50){\mbox{\large $\rho t$}}
\put(-110,-3){\mbox{\large $\rho z$}}
\quad\quad & \quad\quad
\includegraphics[width=0.46 \textwidth]{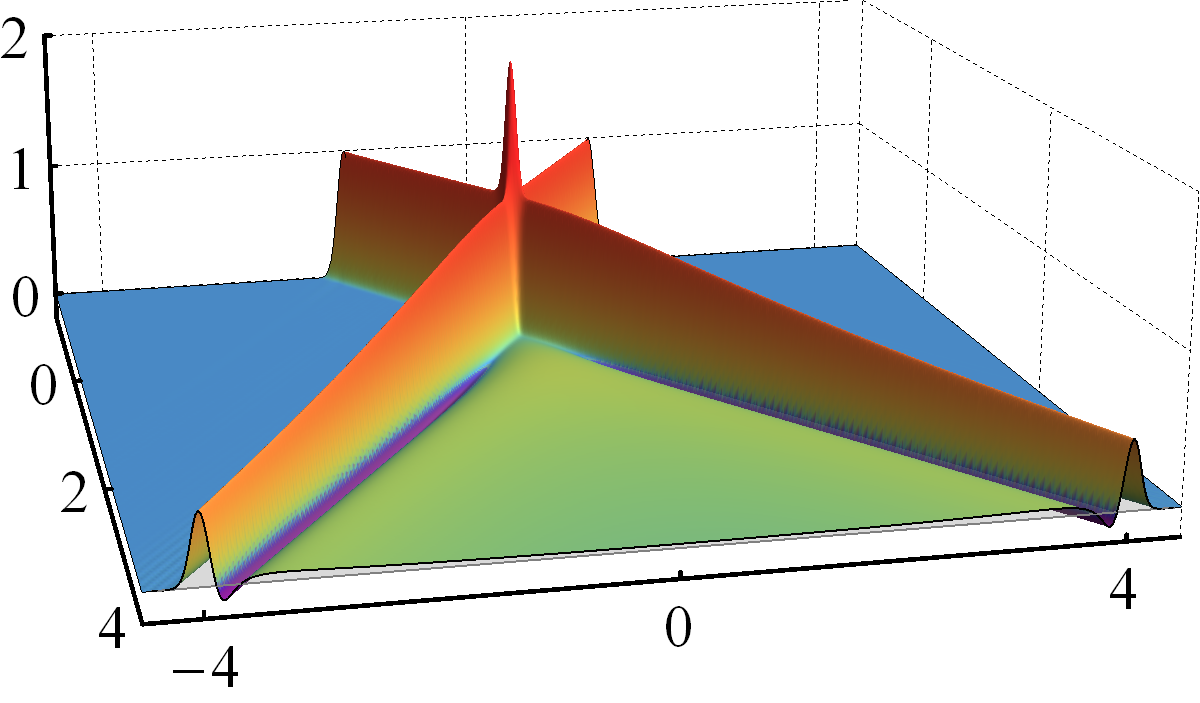}
\put(-167,145){\mbox{\large ${\cal E}/\rho^4$}}
\put(-243,50){\mbox{\large $\rho t$}}
\put(-110,-1){\mbox{\large $\rho z$}}
\\ [5mm]
\includegraphics[width=0.46 \textwidth]{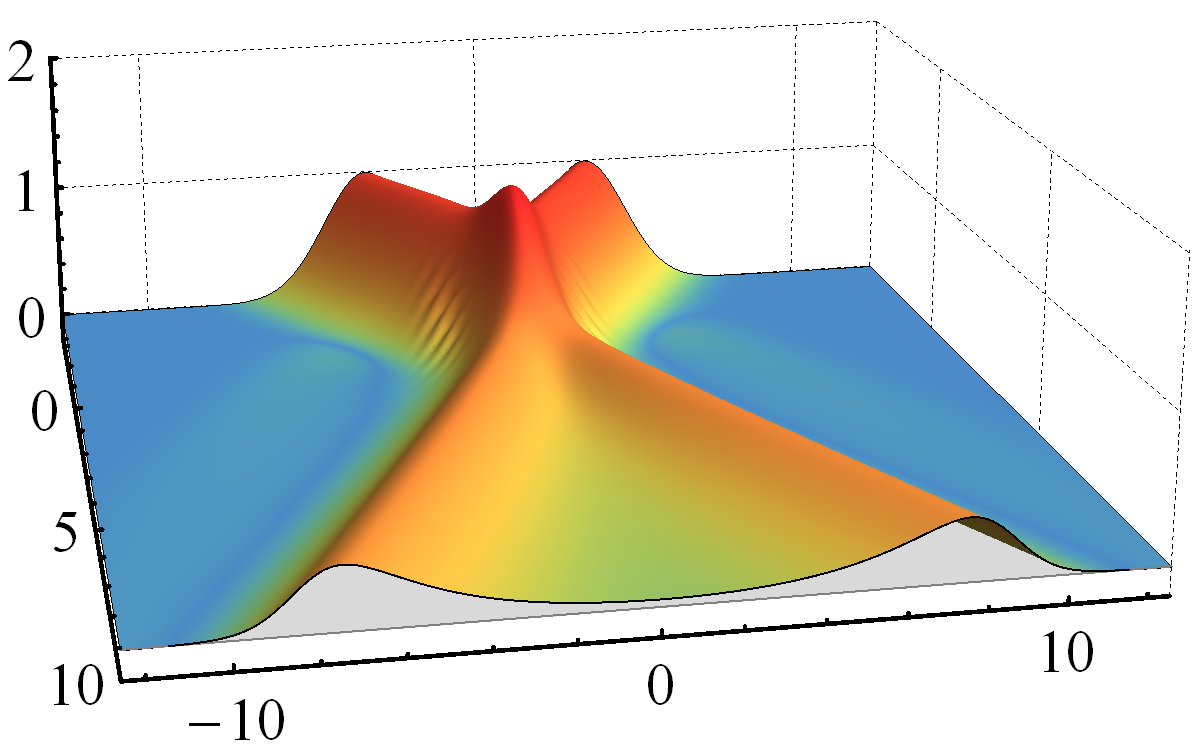}
\put(-167,145){\mbox{\large ${\cal P}_L/\rho^4$}}
\put(-243,50){\mbox{\large $\rho t$}}
\put(-110,-3){\mbox{\large $\rho z$}}
\quad\quad & \quad\quad
\includegraphics[width=0.46 \textwidth]{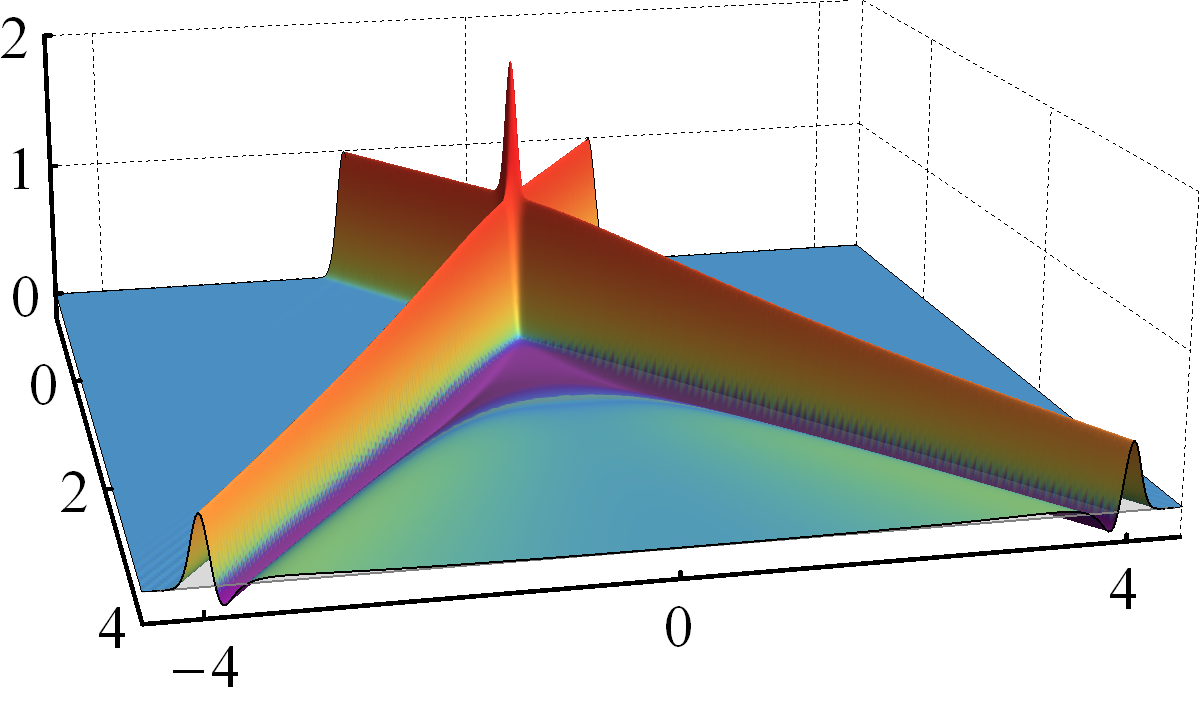}
\put(-167,145){\mbox{\large ${\cal P}_L/\rho^4$}}\put(-243,50){\mbox{\large $\rho t$}}
\put(-110,-3){\mbox{\large $\rho z$}}
\\ [5mm]
\includegraphics[width=0.46 \textwidth]{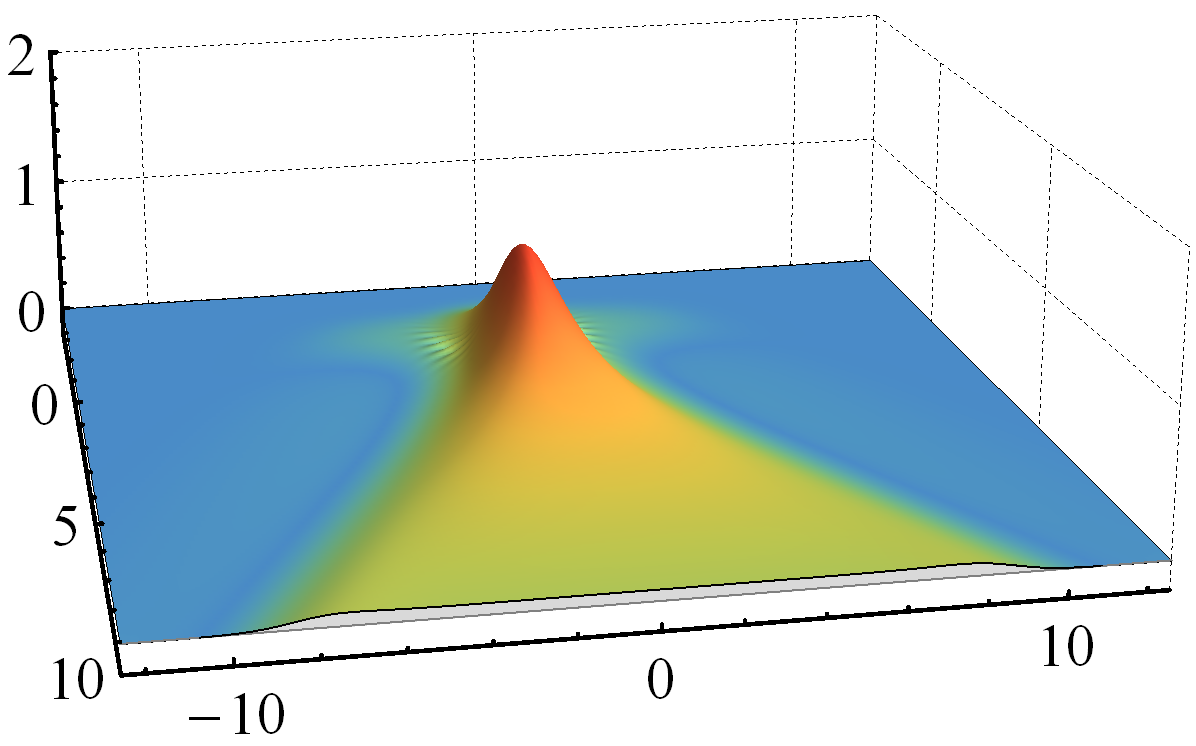} 
\put(-167,145){\mbox{\large ${\cal P}_T/\rho^4$}}
\put(-243,50){\mbox{\large $\rho t$}}
\put(-110,-3){\mbox{\large $\rho z$}}
\quad\quad & \quad\quad
\includegraphics[width=0.46 \textwidth]{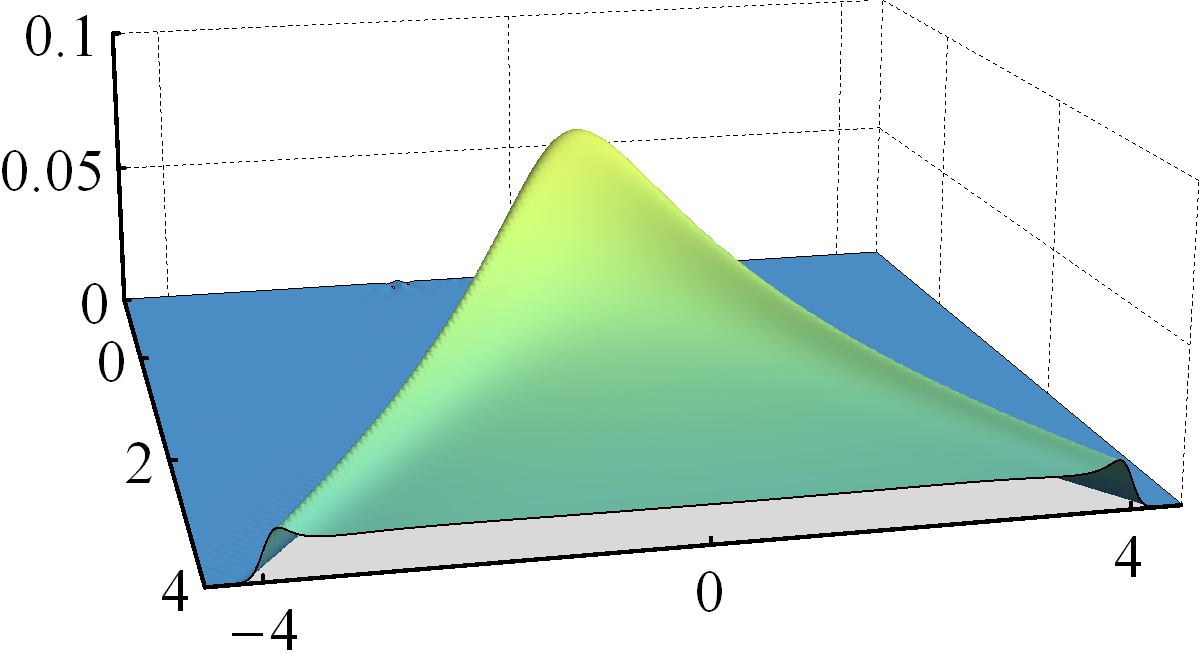}
\put(-167,135){\mbox{\large ${\cal P}_T/\rho^4$}}
\put(-232,50){\mbox{\large $\rho t$}}
\put(-100,-3){\mbox{\large $\rho z$}}
\end{tabular}
\caption{Energy and pressures 
for collisions of thick (top row) and thin  (bottom row) shocks. The grey planes lie at the origin of the vertical axes.
}
\label{EnergyDensity}
\end{figure*} 

We will uncover a cross-over between two qualitatively different dynamical regimes that correspond to a full-stopping scenario for thick shocks, and to a transparency scenario for thin ones. 
Among other things, the two regimes are distinguished by the applicability of hydrodynamics. We will say that hydrodynamics is applicable when the constitutive relations of first-order, viscous hydrodynamics predict $\pl$ in the local rest frame in units of $\eps_\mt{loc} /3$ with a 20\% accuracy, i.e.~when $3 \left| \Delta {\cal P}_{L}^\mt{loc}\right|/\eps_\mt{loc} \leq 0.2$ with $\Delta {\cal P} = {\cal P} - {\cal P}_{\mt{hydro}}$. Tracelessness of the stress tensor then implies that  
$3 \left| \Delta {\cal P}_{T}^\mt{loc}\right|/\eps_\mt{loc} \leq 0.1$.
We define the hydrodynamization time, $\thy$, as the time after which hydrodynamics becomes applicable at $z=0$. Other reasonable definitions include 
$t^\mt{max}_\mt{hyd} =  \thy-\tmax$ and $t^{2w}_\mt{hyd} = \thy+2w$. The former measures hydrodynamization from the time when the energy density achieves its maximum value (see Fig.~\ref{EnergyDensity}). The latter measures hydrodynamization from the time when the two incoming shocks begin to overlap significantly \cite{Chesler:2010bi}. The differences between these definitions are significant for thick shocks but become small for thin shocks. We will also consider another hydrodynamization time, $t_\mt{hyd}^{\cal P}$, defined by the criterion $\left| \Delta {\cal P}_{L}^\mt{loc} \right| /{\cal P}_{L}^\mt{loc} \leq 0.2$. One advantage of $\thy$ over $t_\mt{hyd}^{\cal P}$ is that $\eps_\mt{loc}$ is always non-zero, whereas ${\cal P}_{L}^\mt{loc}$ may vanish.

\noindent
{{\bf 2. A dynamical cross-over.}} Fig.~\ref{EnergyDensity} shows the  energy density and the pressures for thick and thin shock collisions. In the case of $\eps$ and $\pl$ one can see the incoming shocks at the back of the plots, the collision region in the center, and the receding maxima at the front. The incoming shocks are absent in the case of $\pt$, as expected. A simultaneous rescaling of $\rho$ and $w$ that keeps $\rho w$ fixed would change the  overall scales on the axes of these figures but would leave the physics unchanged.
\begin{figure*}[ht]
\begin{center}
\begin{tabular}{cc}
\includegraphics[height=54mm]{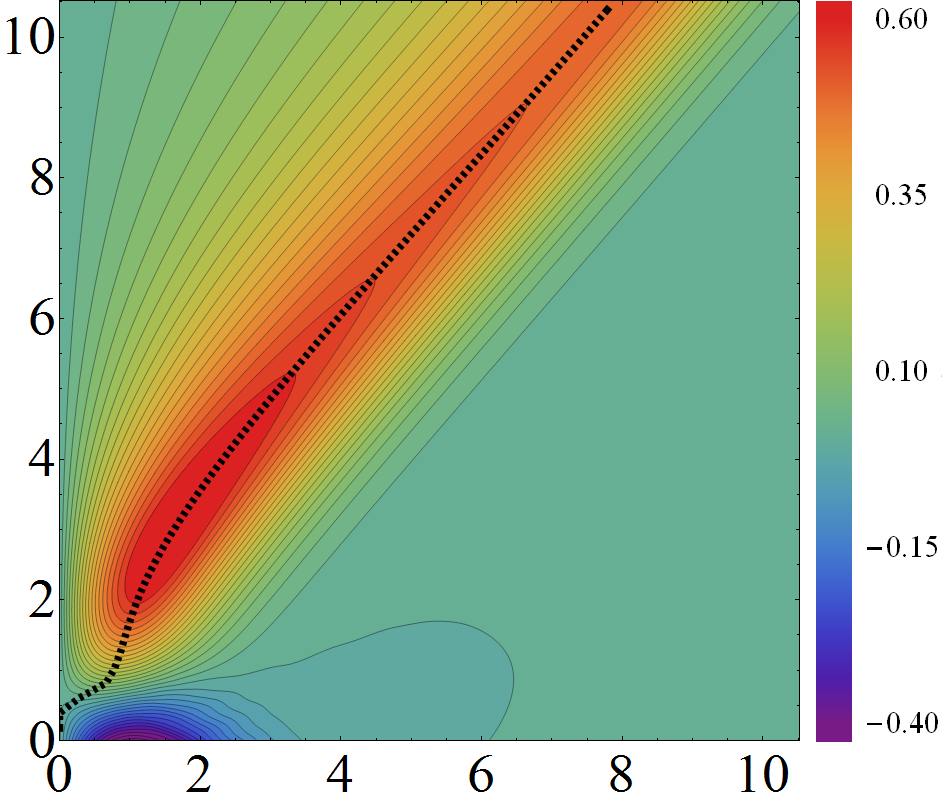} 
\put(-192,75){\rotatebox{0}{\mbox{\large $\rho t$}}}
\put(-105,-10){\rotatebox{0}{\mbox{\large $\rho z$}}}
\put(-107,160){\rotatebox{0}{\mbox{\large $S/\rho^4$}}}
\quad\quad&\quad\quad
\includegraphics[height=54mm]{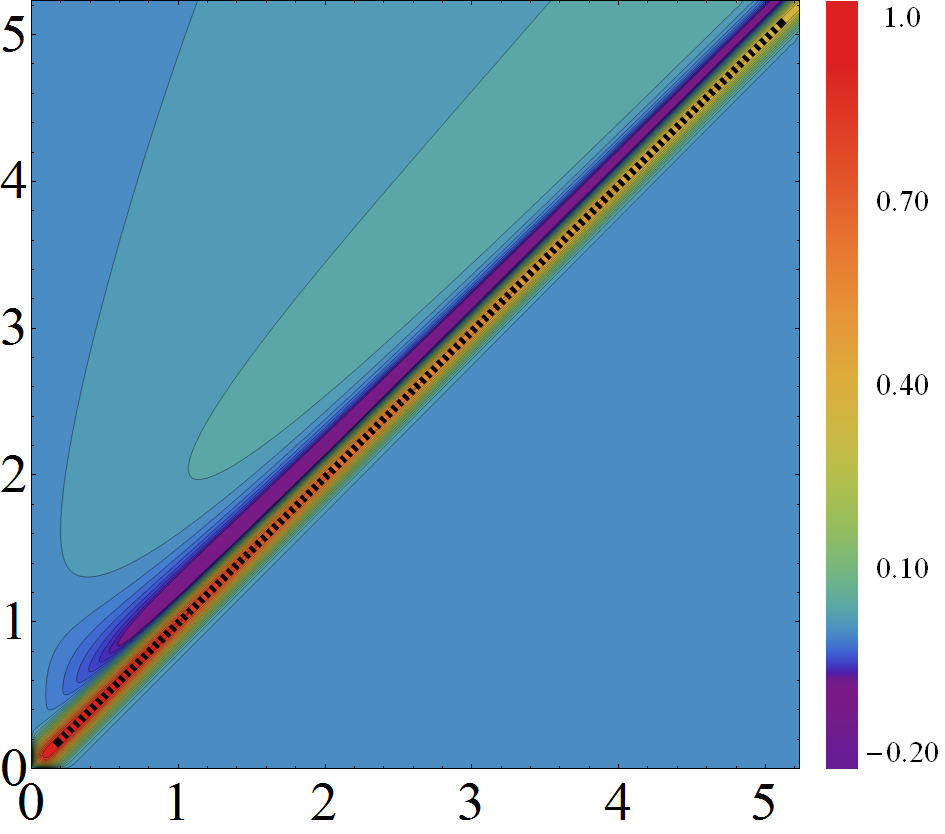} 
\put(-192,75){\rotatebox{0}{\mbox{\large $\rho t$}}}
\put(-105,-10){\rotatebox{0}{\mbox{\large $\rho z$}}}
\put(-107,160){\rotatebox{0}{\mbox{\large $S/\rho^4$}}}
\end{tabular}
\end{center}
\caption{Energy flux for collisions of thick (left) and thin (right) shocks. The dotted curves show the location of the maxima of the flux.}
\label{flux}
\end{figure*} 

\begin{figure*}
\begin{center}
\includegraphics[width=0.75 \textwidth]{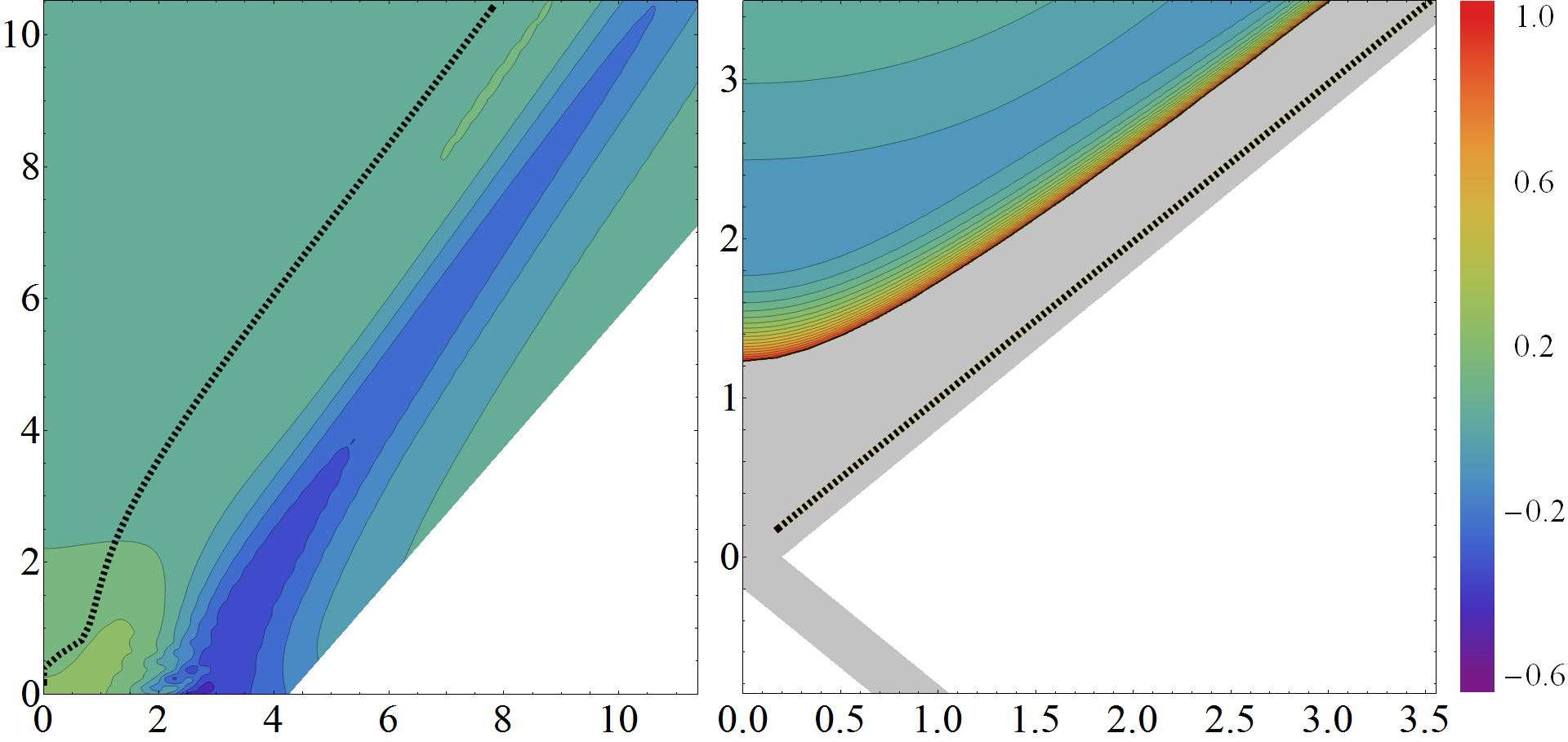} 
\put(-123,-10){\mbox{\large $\rho z$}}
\put(-307,-10){\mbox{\large $\rho z$}}
\put(-398,95){\mbox{\large $\rho t$}}
\end{center}
\caption{$3 \Delta {\cal P}_{L}^\textrm{loc}/\eps_\mt{loc}$ for thick (left) and thin (right) shocks. The white areas indicate the vacuum regions outside the light cone. The grey areas indicate regions where hydrodynamics deviates by more than 100\%. The dotted curves indicate the location of the maxima of the energy flux, as in Fig.~\ref{flux}.}
\label{hydro}
\end{figure*} 

\begin{figure*}
\begin{center}
\begin{tabular}{cc}
\includegraphics[width=68mm, height=44mm]{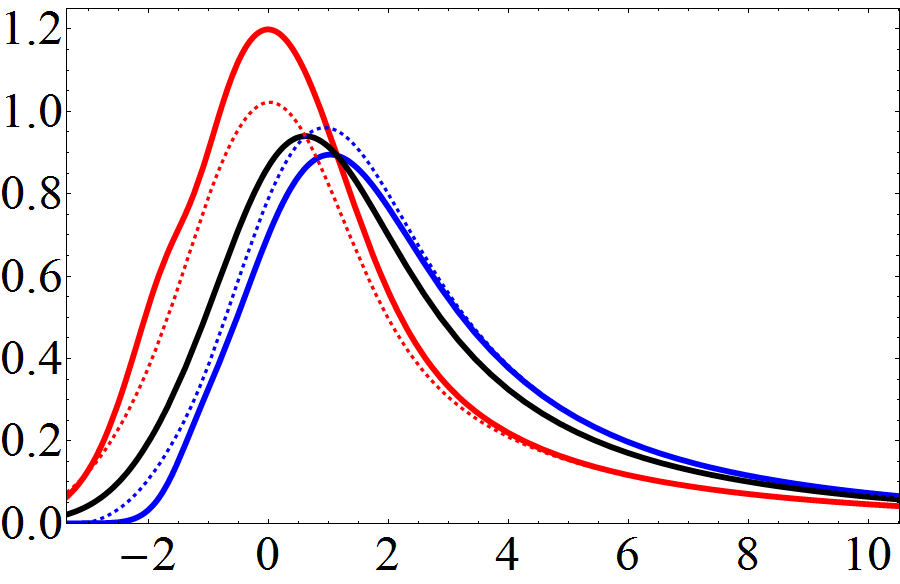} 
\put(-37,112){\fbox{\mbox{\large $\rho z=0$}}}
\put(-95,-9){\mbox{\large $\rho t$}}
\quad\,&\quad\,
\includegraphics[width=68mm, height=45mm]{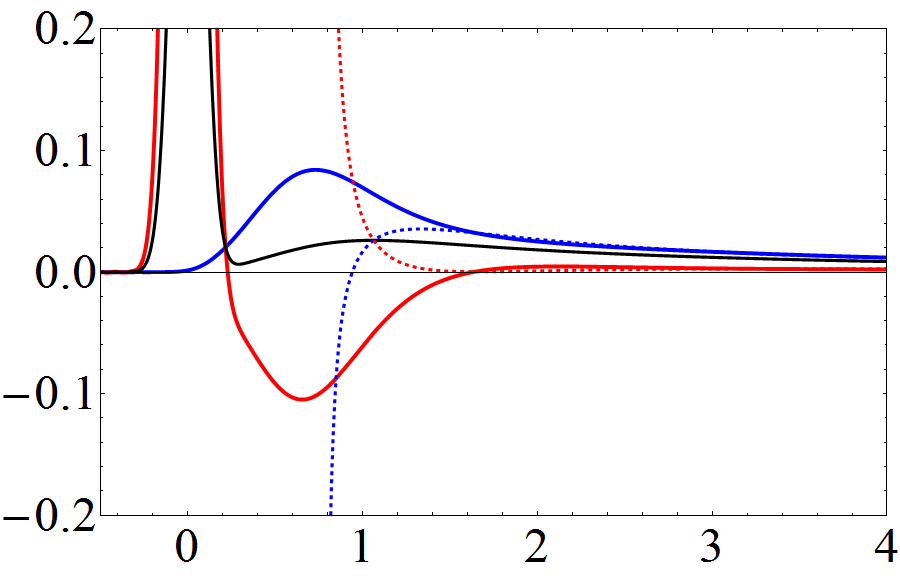} 
\put(-39.5,110.5){\fbox{\mbox{\large $\rho z=0$}}}
\put(-95,-9){\mbox{\large $\rho t$}}
\\
[7mm]
\,
\includegraphics[width=70mm, height=44mm]{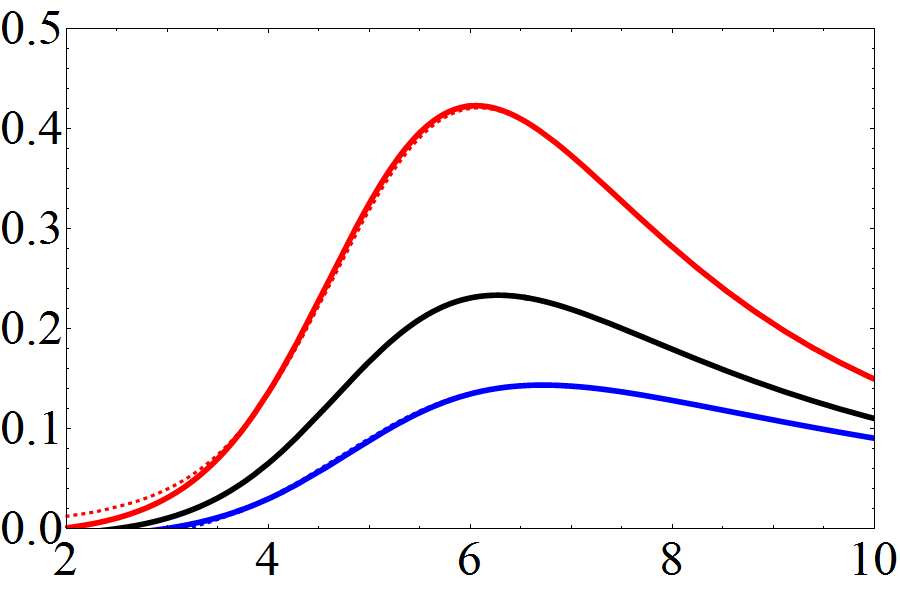}
\put(-42.5,108){\fbox{\mbox{\large $\rho z=4$}}}
\put(-95,-9){\mbox{\large $\rho t$}}
\quad\,&\quad\,
\includegraphics[width=68mm, height=44mm]{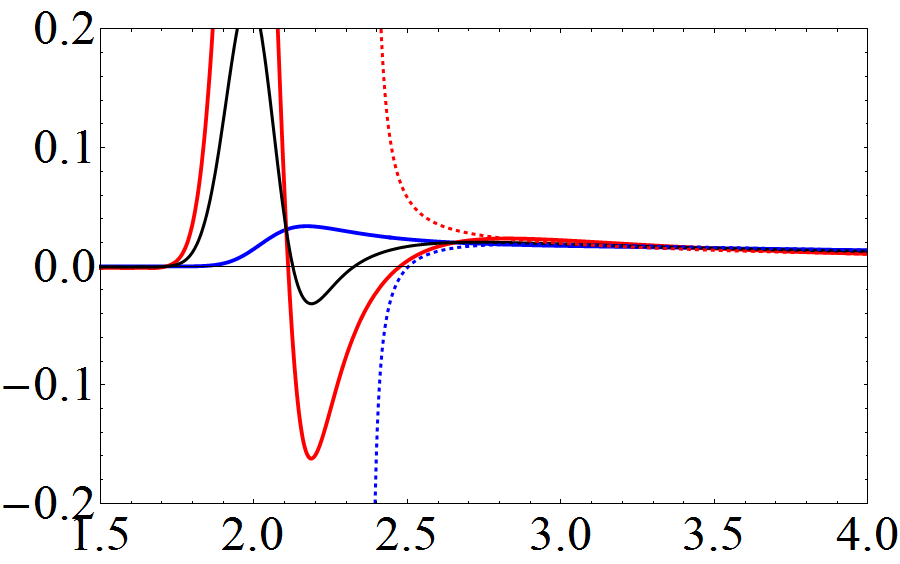}
\put(-43.5,107.7){\fbox{\mbox{\large $\rho z=2$}}}
\put(-95,-9){\mbox{\large $\rho t$}}
\end{tabular}
\end{center}
\caption{Centre and off-center values of $\eps/3 \rho^4$ (black), $\pl/\rho^4$ (red) and $\pt/\rho^4$ (blue) as a function of time for a collision of thick (left column) and thin (right column) shocks. The dotted  curves show the hydrodynamic approximation. 
}
\label{2Dhydro}
\end{figure*} 

The thick shocks illustrate the full-stopping scenario. As the shocks start to interact the energy density gets compressed and `piles up', comes to an almost complete stop, and subsequently explodes hydrodynamically. Indeed, at the time $\rho t_\mt{max} \simeq 0.58$ at which the energy density reaches its maximum in the top-left plot, the energy density profile is very approximately a rescaled version of one of the incoming Gaussians, with about three times its height (see table \ref{table}) and {\small 2/3} its width. At this time, 90\% of the energy is contained in a region of size $\Delta z \simeq 2.4 w$ in which the flow velocity is everywhere $|v|\lesssim 0.1$.
Similarly, the energy flux in this region is less than $10 \%$ of the maximum incoming flux, as illustrated by Fig.~\ref{flux}(left). At late times, the velocity of the receding shocks can be read off from the same figure as the inverse slope of the dotted line. This  is not constant in time, but at late times it reaches a maximum of about $v\simeq 0.88$. The validity of the hydrodynamic description can be seen in Fig.~\ref{hydro}(left) and Fig.~\ref{2Dhydro}(top row). Hydrodynamics becomes applicable even earlier than $\tmax$, and  the region where it is applicable extends from $z=0$ to the location of the receding maxima. This is intuitive since gradients become smaller as $w$ increases. We conclude that the thick-shock collision results in hydrodynamic expansion with initial conditions in which all the velocities are close to zero. This is in close similarity with the Landau model \cite{Landau:1953gs}, which seems to reproduce some aspects of RHIC collisions \cite{Back:2006yw}.

The thin shocks illustrate the transparency scenario. In this case the shocks pass through each other and, although their shape gets altered, they keep moving at $v\simeq 1$, as seen  in  Fig.~\ref{flux}(right). The most dramatic modification in their shape is a region of negative $\eps$ and $\pl$ that trails right behind the receding shocks. While the negative $\eps$ only develops away from the center of the collision, the negative $\pl$ is already present at $z=0$, as shown more clearly in the bottom-left plot of Fig.~\ref{2Dhydro}. These features are compatible with the general principles of Quantum Field Theory \cite{Ford:1999qv}, since the `negative region'  is far from equilibrium and highly localized near a bigger region with positive energy and pressure. In the case of thin shocks, we see from Fig.~\ref{hydro}(right) and Fig.~\ref{2Dhydro}(bottom row) that there is a clear separation between non-hydrodynamic receding maxima and a plasma in between them that is described by hydrodynamics only at sufficiently late times. At sufficiently late times it is also visible from Fig.~\ref{EnergyDensity} that the receding maxima suffer significant attenuation  \cite{unrelated}. We therefore emphasize that our use of the term `transparency' refers to time scales longer than $\thy$ but shorter than the attenuation time. 

Several quantities of interest are given in Table \ref{table}. We see that $\tmax>0$ for thick shocks, whereas for thin shocks $\tmax \simeq 0$, as it would be in the absence of interactions. Similarly, the maximum energy density $\eps_\mt{max}$ is just the sum of the incoming energies for thin shocks, indicating that, unlike for thick shocks, there is no compression or piling up for thin shocks. The minimum energy density $\eps_\mt{min}$ is negative for sufficiently thin shocks, as expected. The fact that $\thy<0$ is negative for thick shocks simply means that hydrodynamics becomes applicable even before the shocks fully overlap. In terms of the criterion $\left| \Delta {\cal P}_{L}^\mt{loc} \right| /{\cal P}_{L}^\mt{loc} \leq 0.2$, hydrodynamics becomes applicable for thick shocks after this full-overlap time but still before the complete stop, i.e.~$0< t_\mt{hyd}^{\cal P} < \tmax$. Roughly speaking, both $\thy$ and $t_\mt{hyd}^{\cal P}$ increase in units of $\rho^{-1}$ or $w$, and decrease in units of  $\mu^{-1}$, as the width decreases. The difference between $\thy^\mt{max}$ and $\thy$ becomes insignificant for thin shocks. As the width decreases, $\thy^{2w}$ first decreases and then increases, the reason being that $\thy^{2w}$ is dominated by $2w$ ($\thy$) for thick (thin) shocks. The  hydrodynamization temperature, $\Thy$, decreases  with decreasing width in units of $\rho$ or $w^{-1}$. In contrast, $\Thy$ is almost constant in units of $\mu$; we will come back to this in Sec.~3. As in other  models \cite{Chesler:2010bi,Chesler:2008hg}, the products $\thy \Thy$ and $\thy^{\cal P} \Thy$ are smaller than unity and fairly constant, which for typical values of $\Thy$ at RHIC and LHC leads to hydrodynamization times (significantly) shorter than 1 fm. The anisotropy $\pt/\pl$ at these times increases as the width decreases, reaching values as large as $\sim 15$. It is remarkable that such strong anisotropies can be well described by first-order hydrodynamics.

\begin{table*}[tdp]
\begin{center}
\begin{tabular}{|c|c|c|c|c|c|c|c|c|c|c|c|c|c|c|c|c|c|c|} 
\hline
\rule{0pt}{3.5ex}
$\frac{\textstyle \rho w}{\textstyle \rho w_\mt{CY}}$ & 
$\rho w$ &
$\mu w$ &
$\rho \tmax$ &
$\frac{\textstyle \mathcal{E}_{\max}}{\textstyle \rho^4}$ & 
$\frac{\textstyle \mathcal{E}_{\min}}{\textstyle \rho^4}$ & 
$\rho \thy$ & 
$\mu \thy$ & 
{$\frac{\displaystyle \thy}{\textstyle w}$} &
$\rho t^\mt{max}_\mt{hyd}$ & 
$\rho t^{2w}_\mt{hyd}$ & 
$\rho t^{\cal P}_\mt{hyd}$ & 
$\frac{\textstyle \Thy}{\textstyle \rho}$ & 
$\frac{\textstyle \Thy}{\textstyle \mu}$ & 
$T_\mt{hyd} w$ & 
$\thy T_\mt{hyd}$ & 
$t_\mt{hyd}^{\cal P} T_\mt{hyd}$ &
$\left. \frac{\textstyle \pt}{\textstyle \pl}\right|_{\thy}$ & 
$\left. \frac{\textstyle \pt}{\textstyle \pl}\right|_{\thy^{\cal P}}$  
\\ [2ex]    
\hline
2 & 1.28 & 1.89 & 0.58 & 2.9 & 0. & -0.053 & -0.078 & -0.041 & -0.63 & 2.5 & 0.34 & 0.44 & 0.30 & 0.56 & -0.02 & 0.15 & 0.54 & 0.70 \\
�1 & 0.64 & 0.75 & 0.13 & 2.3 & 0. & 1.2 & 1.5 & 2.0 & 1.1 & 2.5 & 1.6 & 0.36 & 0.31 & 0.23 & 0.45 & 0.58 & 3.2  & 3.1 \\
�{\scriptsize $1/2$} & 0.32 & 0.30 & 0.03 & 2.0 & 0. & 1.1 & 1.0 & 3.4 & 1.0 & 1.7 & 2.1 & 0.29 & 0.31 & 0.093 & 0.32 & 0.61 & 6.2 & 3.4 \\
�{\scriptsize $1/4$} & 0.16 & 0.12 & 0. & 2.0 & 0. & 1.2 & 0.88 & 7.5 & 1.2 & 1.5 & 2.2 & 0.22 & 0.30 & 0.035 & 0.27 & 0.48 & 12.  & 4.3 \\
�{\scriptsize $3/16$} & 0.12 & 0.08 & 0. & 2.0 & -0.01 & 1.3 & 0.88 & 11. & 1.3 & 1.6 & 2.4 & 0.20 & 0.30 & 0.024 & 0.27 & 0.49 & 11. & 4.9 \\
�{\scriptsize $1/8$} & 0.08 & 0.05 & 0. & 2.0 & -0.1 & 1.5 & 0.87 & 19. & 1.5 & 1.7 & 2.4 & 0.17 & 0.30 & 0.014 & 0.26 & 0.42 & 15. & 4.6 \\ 
\hline
\end{tabular}
\end{center}
\vspace{-2.8mm}
\caption{Numerical values of several quantities of interest.}
\vspace{-0.5mm}
\label{table}
\end{table*}
\begin{figure*}
\begin{center}
\begin{tabular}{cc}
\includegraphics[width=0.35 \textwidth,height=40mm]{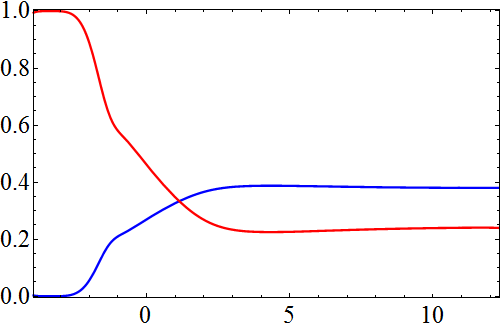} 
\put(-85,-8){\mbox{\large $\rho t$}}
\quad\, & \quad\,
\includegraphics[width=0.37 \textwidth,height=40mm]{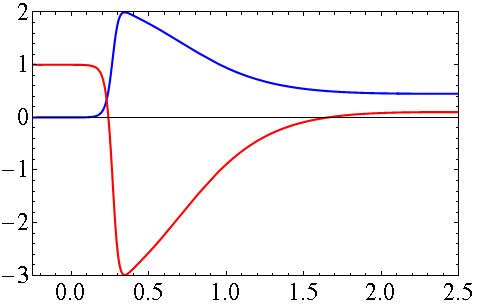} 
\put(-85,-8){\mbox{\large $\rho t$}}
\end{tabular}
\end{center}
\caption{Ratios ${\cal P}_L/\eps$ (red) and ${\cal P}_T/\eps$ (blue) at $z=0$ for thick (left) and thin (right) shocks.}
\label{ratios}
\end{figure*} 
\begin{figure*}
\begin{tabular}{cc}
\includegraphics[width=0.4 \textwidth]{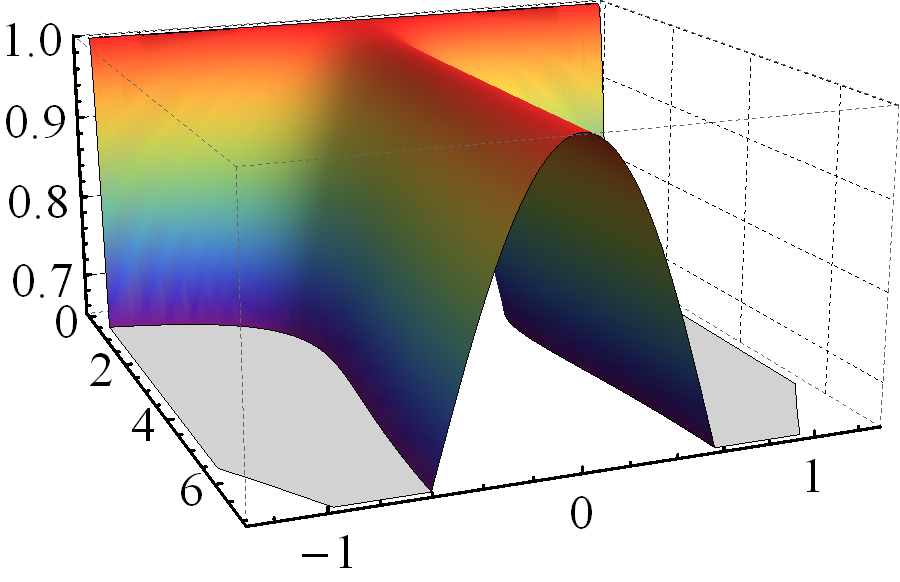} 
\put(-198,35){\mbox{\large $\rho \tau$}}
\put(-70,2){\mbox{\large $\eta$}}
\put(-160,140){$\eps_\mt{loc}(\tau,\eta)/\eps_\mt{loc} (\tau,\eta=0)$}
\quad\,\,\,\,&\,\,\,\, \quad
\includegraphics[width=0.4 \textwidth]{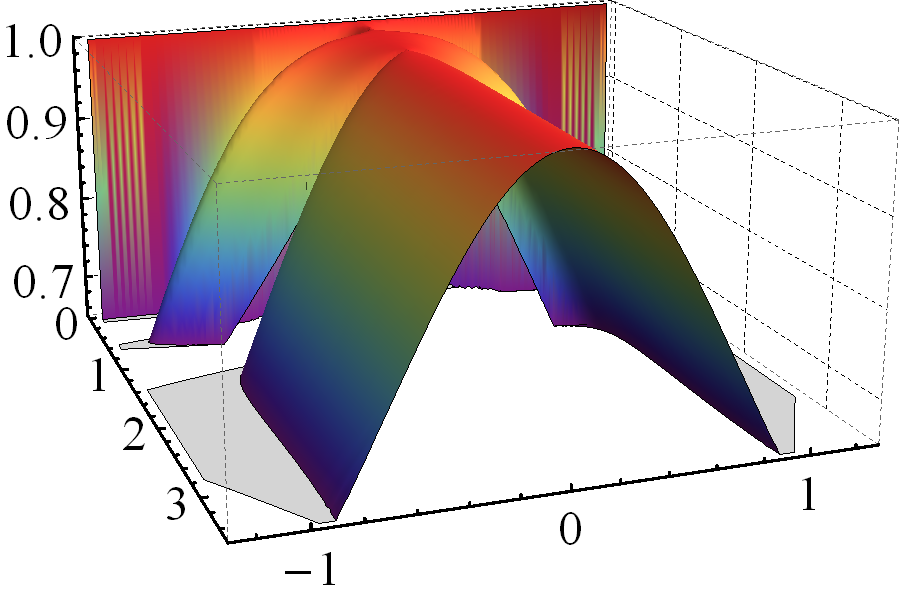} 
\put(-198,35){\mbox{\large $\rho \tau$}}
\put(-70,2){\mbox{\large $\eta$}}
\put(-160,140){$\eps_\mt{loc}(\tau,\eta)/\eps_\mt{loc} (\tau,\eta=0)$}
\end{tabular}
\caption{Energy density in the local rest frame around mid-rapidity as a function of spacetime rapidity $\eta$ and proper time $\tau$ for thick  (left) and thin  (right) shocks. In the latter case we have excluded from the plot the region in which the local rest frame is not defined because $2 |{\cal S}| > |\eps + \pl |$.
}
\label{rapidity}
\end{figure*} 

\noindent
{{\bf 3. Discussion.}} 
The crossover can be heuristically understood on the gravity side. Since each of the colliding shock waves is a normalizable solution in the bulk, the metric near the AdS boundary is a small deviation from  AdS. Consequently, the gravitational evolution is linear near the boundary for some time $\tlin$. The deviation becomes of order one at $u \sim \rho^{-1}$, with $u$ the usual Fefferman-Graham holographic coordinate. At this depth gravity becomes strong and the evolution is non-linear. This non-linearity takes $\tlin \sim u \sim \rho^{-1}$ to propagate to the boundary. If $w \ll \tlin$, i.e.~if $\rho w \ll 1$, there is a clear separation between the linear and the non-linear regimes. For thin shocks, this is illustrated by e.g.~Fig.~\ref{2Dhydro}(bottom-left), where the energy density exhibits two maxima around $\rho t \sim 0$ and $\rho t \sim 1$. The former corresponds to the two shocks passing through each other; the latter corresponds to the arrival to the boundary of the non-linear pulse from the bulk. In this sense the pulse is responsible for the `creation' of the plasma in between the thin receding shocks. In contrast, for thick shocks $\rho w \gg 1$, meaning that $\tlin \ll w$. In this case the pulse reaches the boundary  before the shocks have passed through each other and essentially all the evolution is non-linear. 

This analysis  suggests that we have identified all the qualitatively different dynamical regimes. Presumably we have also considered values of $\rho w$ sufficiently representative of the asymptotic regimes $\rho w \gg 1$ and $\rho w \ll 1$. For thick shocks this is suggested by the fact that they come very close to a complete stop and subsequently evolve  hydrodynamically. For thin shocks this is suggested by comparison of Fig.~\ref{ratios}(right) with \cite{Grumiller:2008va}. This reference studied the delta-function limit $\rho\to \infty, w \to 0$ with $\mu$ fixed and found that the pressure/energy ratios are 
${\cal P}_L/\eps = -3$ and ${\cal P}_T/\eps=2$ at $t\to 0^\plus$. 
Fig.~\ref{ratios}(right) shows that these are also the extremum values attained by our thin shocks.

The scaling $\Thy \simeq 0.3 \mu$ shown in Table \ref{table} is remarkable. First, it relates $\Thy$ to the same property of the initial state for collisions that reach hydrodynamization through qualitatively  different dynamics. Second, it shows that $\Thy$ is independent of how the initial transverse energy density is distributed along the longitudinal direction, which is reminiscent of the scaling with the number of participants observed in HIC. 
In combination with the $\thy \Thy$ column, this scaling implies that $\thy \Thy^3 \sim \mu^2 \sim s_\mt{coll}^{1/3}$. The product $\thy \Thy^3$ may be taken as a crude proxy for the multiplicity per unit rapidity at mid-rapidity in our model, since it measures the entropy density per unit rapidity and per unit transverse area at $\thy$.
The {\small 1/3} exponent in $s_\mt{coll}$ is a factor of 2 larger than the experimental value \cite{Aamodt:2010pb}, which might be due to the fact that our system is strongly coupled at all scales. 

Our results dispel two possible preconceptions. First, they show that infinite coupling in the CFT need not lead to any significant stopping and is compatible with receding shocks moving at the speed of light. Second, they illustrate that the latter property does not necessarily lead to  boost invariance at mid-rapidity. This is clearly seen in Fig.~\ref{rapidity}, where we have changed to proper-time and spacetime-rapidity coordinates. The `tubes' at late times show that the local energy density at mid-rapidity is not rapidity-independent but  has a Gaussian profile. Yet, it is interesting  that the width of this Gaussian increases as $w$ decreases, in agreement with general expectations.

\noindent
{{\bf Acknowledgements.}} We thank G.~Arutyunov, P.~Chesler, J.~Garriga, R.~Janik, T.~Peitzmann, K.~Rajagopal, P.~Romatschke, R.~Snellings and D.~Teaney for discussons. JCS and DM acknowledge financial support from grants FPA2010-20807 and CPAN CSD 2007-00042 Consolider-Ingenio 2010. JCS is further supported by a RyC fellowship and by grants 2009SGR502 and FP7-PEOPLE-2012-GIG-333786. MPH is supported by the Netherlands Organization for Scientific Research under the NWO Veni scheme (UvA) and by the National Science Centre under Grant No. 2012/07/B/ST2/03794 (NCNR). DM is also supported by grants ERC StG HoloLHC-306605 and 2009SGR168. WS is supported by a Utrecht University Foundations of Science grant. We used M. Headrick's excellent Mathematica package \href{http://people.brandeis.edu/~headrick/Mathematica/index.html}{\tt diffgeo.m}.




\begin{thebibliography}{99}
\bibitem{transverse}
Note that the longitudinal structure and the transverse structure of our incoming shocks are a priori uncorrelated in the sense that $\mu$ and $\omega$ can be chosen independently. To make contact with a real collision one may identify $\omega\sim L/\gamma$, with $L$ the width of the nucleus in the rest frame, and $\rho^4 \sim \rho_\mt{rest}^4 \gamma^2$, with $\rho_\mt{rest}^4$ the nucleus density in the rest frame. Then one finds $\mu^3 \sim \mu_\mt{rest}^3 \gamma$, indicating that the shocks become denser in the transverse directions as the gamma factor of the collision increases. The saturation scale and its evolution in the holographic direction for our incoming shocks were studied in \cite{saturation}.
\bibitem{regulator}
We employ as a regulator a background thermal bath with $\eps$ between $0.015 \rho^4$ and $0.075 \rho^4$. In particular, this means that a black brane horizon is present at all times. For numerical convenience, we use coordinates in the AdS bulk such that the apparent horizon lies at a constant position along the holographic radial coordinate.
\bibitem{unrelated}  
We have checked that this is unrelated to our regulator by comparing simulations with two different regulators.
\end{thebibliography}

\end{document}